\documentclass[aip,letterpaper,11pt,nofootinbib]{revtex4}

\usepackage{amsmath,mathtools}    
\usepackage{amssymb}    
\usepackage{mathrsfs}
\usepackage{graphicx}   
\usepackage{float}
\usepackage{chemfig}
\usepackage{array}
\usepackage{mdframed}
\usepackage{multirow}
\usepackage{pdflscape}
\usepackage{verbatim}   
\usepackage{subfigure}  
\usepackage{framed}

\newcommand*\colvec[3][]{
    \begin{pmatrix}\ifx\relax#1\relax\else#1\\\fi#2\\#3\end{pmatrix}
}

\def\app#1#2{%
  \mathrel{%
    \setbox0=\hbox{$#1\sim$}%
    \setbox2=\hbox{%
      \rlap{\hbox{$#1\propto$}}%
      \lower1.1\ht0\box0%
    }%
    \raise0.25\ht2\box2%
  }%
}

\def\kisc{k_{isc}}
\def\krisc{k_{risc}}
\def\e{\operatorname{e}}
\def\diffs{\left(k_1-k_3+\kisc-\krisc\right)}
\def\kd{k_{\Delta}}
\def\dE{\Delta E_{S-T}}

\providecommand{\EE}[1]{\ensuremath{\times 10^{#1}}}

\begin{document}

\title{Rates and singlet/triplet ratios from TADF transients}
\author{Mitchell C. Nelson}
\email[Correspondence:\ ]{drmcnelsonm@gmail.com}

\date{15 March 2016}

\begin{abstract}
  Thermally activated delayed fluorescence has been reported in a number of OLED emitter materials engineered to have low singlet-triplet energy gaps.
  Here we derive closed solutions for steady state and transient behaviors, and apply these results to obtain the singlet and triplet relaxation, forward and reverse crossing rates and the gap energy and reverse crossing prefactor from delayed and prompt relaxation rates measured over a series of temperatures.
  The primary rates are then used to calculate the fluorescent/phosphorescent ratio and the singlet/triplet population ratio.
  The method avoids the need for gated yield measurements.
  Good fits are obtained using previously published data for 4CzIPN and m-MTDATA:t-Bu-PBD and the results appear to be consistent with reported quantities were available, and with reported behaviors of OLEDS that use these materials.
\end{abstract}

\maketitle

\section{Introduction}

Thermally activated fluorescence (TADF) is seen as a strategy for improving OLED efficiency by harvesting excitation energy from slowly relaxing phosphorescent states via reverse intersystem crossing (RISC) to faster relaxing, more efficient singlet states.
\cite{Endo2011,Goushi2012,Uoyama2012,Dias2013,Niwa2014,Chen2015,Bergmann2016}
Materials engineered to implement TADF have been reported to achieve as high as 100\% internal quantum efficiency with large reverse crossing rates and low singlet-triplet energy gaps.\cite{Uoyama2012,Dias2013,Chen2015}
Yet, singlet/triplet ratios and relaxation rates are determined by at least four parameters comprising the individual relaxation rates and intersystem crossing rates.
Direct observation of these rates has been considered difficult.\cite{Bergmann2016}

Singlet-triplet energy gaps have been obtained from the temperature dependence of reverse crossing rates
where the reverse rates are calculated from prompt and delayed yields and decay rates and a supplied value for the forward crossing rate.\cite{Kirchhoff1983,Goushi2012}
In this work we describe a method for finding the singlet and triplet relaxation rate constants, the forward and reverse intersystem crossing rate constants, and the singlet-triplet energy gap and prefactor governing the reverse intersystem crossing rate constant, using only the temperature dependent decay rates.
The ratio of fluorescence to phosphorescence, and the ratio of singlets to triplets,
in steady state electroluminescence and phototransients, can then be calculated from the four primary rate constants.

In Section~\ref{sec:tadf}, we model the system in a pair of linear differential equations which we solve to obtain the steady state behavior, and the time dependent behavior following from a short photoexcitation into the singlet.
Then in Section~\ref{sec:analysis}, we apply our model to re-analyze published TADF data for two materials to recover their primary relaxation and crossing rates and the reversing crossing parameters.
The results are found to be in agreement with reported values where available and consistent with observed behaviors.

\section{\label{sec:tadf}TADF}
We consider behaviors in two regimes, continuously driven systems as in electroluminescence, and transient luminescence as follows from photoexcitation by a subnanosecond laser pulse.
Closed form solutions are found for both regimes in terms of four primary rates: relaxation from the singlet and from the triplet, and forward and reverse intersystem crossing.
Temperature dependence, scaled by the singlet-triplet gap energy, enters through the thermally activated reverse crossing rate and provides a route to extract the primary parameters from empirical decay rates.

\begin{figure}[h!]
\centering
\includegraphics[scale=1.0]{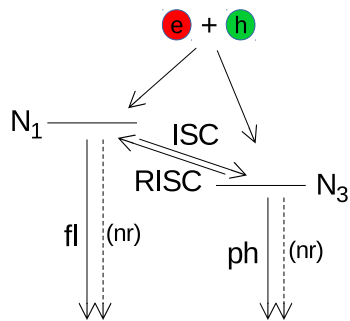}
\caption{\label{fig:electroluminescence}Electroluminescence where charge carrier recombination forms a mix of singlets ($N_1$) and triplets ($N_3$), with fluorescent (fl) and nonradiative (nr) relaxation from the singlets, phosphorescent (ph) and nonradiative relaxation from the triplets, and intersystem crossing between singlets and triplets in the forward (ISC) and reverse (RISC) directions.}
\end{figure}

Electroluminescence occurs when charge carriers (electrons and holes) are combined onto a site where they form an excited state that can decay and emit light.\cite{GasparBook2015}
Charge carriers arrive with random spin and so a mixed population of singlet and triplet excited states is formed, as illustrated in FIG.~\ref{fig:electroluminescence}.

Rate equations for our empirical case can be written as\cite{Peng2012}
\begin{align}
  \dot n &= \frac{\gamma}{eV} I - k_{eh} n^2 N_0 \\
  \dot N_1 &= \eta_{S/T} k_{eh} n^2 N_0 - ( k_1 + \kisc ) N_1 + \krisc N_3 \\
  \dot N_3 &= (1 - \eta_{S/T}) k_{eh} n^2 N_0 - ( k_3 + \krisc ) N_3 + \kisc N_1
\end{align}
where $\gamma I$ is the recombination current,
$e$ is the charge on an electron,
$V$ is the effective volume,
$k_{eh}$ is the recombination rate constant,
$n$ is the free charge carrier density,
$N_0$ is the ground state population (per unit volume),
$N_1$ is the singlet excited state population,
$N_3$ is the triplet excited state population,
$\eta_{S/T}$ is the fraction of singlets produced by recombination,
$k_1$ is the singlet relaxation rate with
$k_1 = k_{fl} + k_{1,nr}$ where $k_{fl}$ is rate of radiative relaxation from the singlet (fluorescence) and $k_{1,nr}$ is the rate of nonradiative 1st order relaxation from the singlet,
and similarly,
$k_3$ is the relaxation rate from the triplet with
$k_3 = k_{ph} + k_{3,nr}$ where $k_{ph}$ is rate of radiative relaxation from the triplet (phosphorescence) and $k_{3,nr}$ is the rate of nonradiative 1st order relaxation from the triplet,
$\kisc$ is the forward intersystem crossing rate
and $\krisc$ is reverse intersystem crossing rate.
For the present analysis we consider a simple system at low power where higher order losses can be omitted.\footnote{Self- and charge-quenching losses in three level systems with crossings are discussed in detail in \cite{Nelson2016}.}

\begin{figure}[h!]
\centering
\includegraphics[scale=1.0]{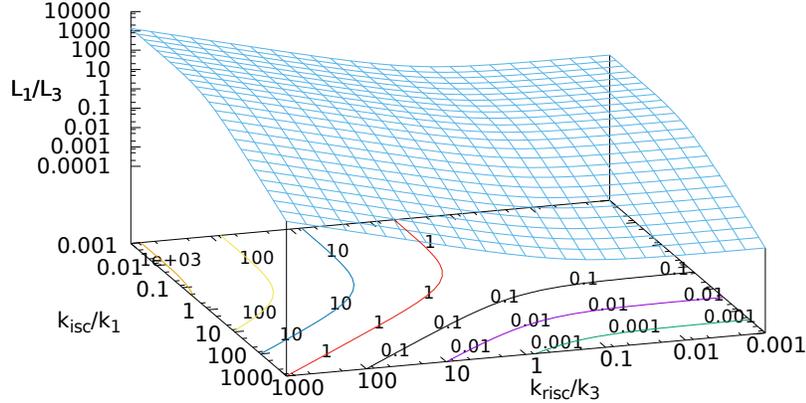}
\caption{\label{fig:outputratio}Fluorescent/phosphorescent output ratio in steady state, as a function of forward and reverse crossing rates in dimensionless units scaled by the spontaneous relaxation rates per equation \eqref{eq:steadyratio}.}
\end{figure}

In steady state, the ratio of the singlet and triplet populations is obtained as
\begin{equation}
  \frac{N_1}{N_3} = \frac{\eta_{S/T} + k_{risc} / k_3}{(1-\eta_{S/T}) + k_{isc} / k_1}\frac{k_3}{k_1}
\end{equation}
and the ratio of fluorescent to phosphorescent output is
\begin{equation}
  \label{eq:steadyratio}
  \frac{L_1}{L_3} = \frac{\eta_{S/T} + k_{risc} / k_3}{(1-\eta_{S/T}) + k_{isc} / k_1}\ \frac{\phi_1\ \chi_1}{\phi_3\ \chi_3}
\end{equation}
where $L_1 = k_{fl} N_1$ is the fluorescent output, $L_3 = k_{ph} N_3$ is the phosphorescent output, $\phi_1 = k_{fl}/k_1$ and $\phi_3 = k_{ph}/k_3$ are the radiative quantum yields for fluorescence and phosphorescence and $\chi_1$ and $\chi_3$ are the fraction of each that escape the device.
In FIG.~\ref{fig:outputratio} the output ratio $L_1/L_3$ is shown as a function of $\kisc/k_1$ and $\krisc/k_3$.
Fluorescence is stronger than phosphorescence in the quadrant where $\krisc/k_3 > 1$ and $\kisc/k_1 <1$.

Temperature dependence enters the fluorescent and singlet ratios through $\krisc$ which is usually thermally activated and follows an Arrhenius law,\cite{Rohatgi1978}
\begin{equation}
  \label{eq:krisc}
  \krisc = \Phi \e^{-\dE/k_B T}
\end{equation}
where $dE$ is the singlet-triplet energy gap, $k_B$ is Boltzmann's constant, T is the Kelvin temperature and the prefactor $\Phi$ is related to spin-orbit coupling mechanisms.\cite{Lawetz1972}
A change in temperature is then equivalent to moving along the surface in FIG.~\ref{fig:outputratio}, in the direction indicated by the axis labeled $\krisc/k_3$.

\begin{figure}[h!]
\centering
\includegraphics[scale=1.0]{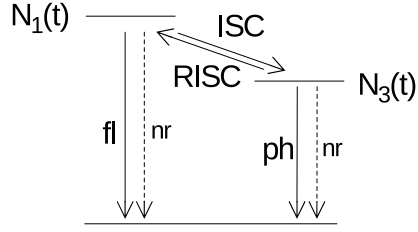}
\caption{\label{fig:photoluminescence}Relaxation and intersystem crossing following photoexcitation, with fluorescent (fl) and nonradiative (nr) relaxation from the singlet and phosphorescent (ph) and nonradiative relaxation from the triplet, and forward (ISC)  and reverse (RISC) intersystem crossings.}
\end{figure}

Transient photoluminescence following sub-nanosecond laser excitation has been used to study several TADF materials.\cite{Goushi2012,Uoyama2012}
As we will show here, two new rate constants emerge in the transient relaxation from a pulsed input in these systems.
The temperature dependence of the emergent rate constants provides an opportunity to obtain the four primary rate constants along with singlet-triplet energy gap and prefactor discussed above.
We begin by finding the closed solution to the relaxation problem.
For our purposes we are interested in the decay following from an initial state $N_1(0), N_3(0)$, as depicted in FIG.~\ref{fig:photoluminescence}.

We write rate equations for the system as
\begin{align}
  \dot N_1 = - ( k_1 + \kisc ) N_1 + \krisc N_3 \label{sdecay}\\
  \dot N_3 = - ( k_3 + \krisc ) N_3 + \kisc N_1 \label{tdecay}
\end{align}
which we will solve subject to initial conditions corresponding to the moment just following a fast pulse from a laser. We present the solution in some detail to illuminate an important behavior of the system.

The rate equations can be written as  $\dot{\vec{\rm N}} = {\bf A} {\vec{\rm N}}$, where
\begin{equation}
  \label{twobytwomatrix}
  {\bf A} =
  \left[ \begin{array}{ccc}
    - ( k_1 + \kisc) & \krisc \\
    \kisc &  - ( k_3 + \krisc)
    \end{array}\right]
\end{equation}
We assume $\vec{N} = \vec{a} {\rm e}^{rt}$ are solutions of the system and find $r$ and $\vec{a}$ as eigenvalues and eigenvectors of $\left|{\bf A} - {\bf I}\ r\right| = 0$.
We thus have two decay constants, given by
\begin{equation}
  r = \frac{1}{2}\left[Tr({\bf A}) \pm \left( {\rm Tr}({\bf A})^2 - 4\ {\rm Det}{\bf A}\right)^{1/2}\right]
\end{equation}
Approximating the square root, we obtain the emergent rate constants,
\begin{align}
  \label{eq:r1}
  r_1 \approx -\left( k_1 + \kisc + \frac{\kisc \krisc}{k_1 - k_3 + \kisc - \krisc}\right)\\[1em]
  \label{eq:r2}
  r_2\approx -\left( k_3 + \krisc - \frac{\kisc \krisc}{k_1 - k_3 + \kisc - \krisc}\right)
\end{align}
The last term can be small so the fast rate constant $r_1$ is less sensitive to changes in $\krisc$ compared to the slow rate constant $r_2$ which increases directly with an increase in $\krisc$.
Eigenvectors can be found as
\begin{equation}
  {\bf a_n} = \colvec{\krisc}{k_1+\kisc+r_n}
\end{equation}
or alternatively,
\begin{equation}
  {\bf a_n} = \colvec{k_3+\krisc+r_n}{\kisc}
\end{equation}
General solutions, using either set of eigenvectors, are
\begin{equation}
  \label{gensol}
  \colvec{N_1}{N_3} = C_1 {\bf a_1 } {\rm e}^{r_1 t} + C_2 {\bf a_2 } {\rm e}^{r_2 t}
\end{equation}
where $C_1$ and $C_2$ are chosen to satisfy initial conditions.  The two emergent rate constants are thus shared by the singlet and the triplet.

Following excitation with a fast laser pulse at or above the singlet absorption, we expect an initial configuration $N_1(0) = 1, N_3(0) = 0$. Choosing the first option for our eigenvectors and applying the initial condition on the triplet we obtain
\begin{equation}
  \frac{C_1}{C_2} = 1 + \frac{\kd^2}{\kisc\krisc}
\end{equation}
where $\kd = \diffs$.
Substituting this into \eqref{gensol}, we obtain
\begin{align}
  \frac{N_1}{C_2} &=
  \left( \krisc + \frac{\kd^2}{\kisc}\right) \e^{r_1 t}
  + \krisc \e^{r_2 t}\\[1em]
  \frac{N_3}{C_2} &=
  \left[ \kd + \frac{\kisc\krisc}{\kd}\right]\left( \e^{r_2 t} - \e^{r_1 t} \right)
\end{align}
During the time $t < 1/r_1$, the triplet population ramps up while the singlet decays, both at $r_1$, and then at some time $t >> 1/r_1$ both the singlet and triplet decay slowly at the shared rate $r_2$.
The delayed luminescence therefore has a constant ratio of fluorescence and phosphorescence,
\begin{equation}
  \label{eq:tranratio}
  \frac{L_3}{L_1} \approx \left( \frac{\kisc}{\kd} + \frac{\kd}{\krisc} \right) \frac{k_3}{k_1}
\end{equation}
(apart from a factor of  $\phi_1\chi_1/\phi_3\chi_3$).

With a minor simplification, the prompt luminescence ($L_1 + L_3$) at sufficiently small $t$ (compared to $1/r_1$) can be written as,
\begin{equation}
  \label{eq:promptluminescence}
  L_p = \left[ \left(\frac{\kd}{\kisc}k_1 - k_3\right) \kd + \left(k_1-\frac{\kisc}{\kd}\right) \krisc \right] C_2 \e^{-r_1 t}
\end{equation}
and at sufficiently large $t$, the delayed luminescence can be written as,
\begin{equation}
  \label{eq:delayedluminescence}
  L_d = \left[k_3 \kd + \left( k_1 + k_3 \frac{\kisc}{\kd}\right) \krisc \right] C_2 \e^{-r_2 t}
\end{equation}
Thus the rate constants $r_1$ and $r_2$ are readily available by simply supplying a short laser pulse to excite the system into the singlet, and then measuring the decay lifetimes within a suitably prompt time interval and after some suitably delayed time interval.

It might be noted on examining equation \eqref{eq:r1} and equation \eqref{eq:r2}, that
\begin{equation}
  r_1 + r_2 = - ( k_1 + k_3 + \kisc + \krisc )
\end{equation}
Measuring $r_1$ and $r_2$ at several temperatures and then fitting the quantity $r_1 + r_2$, to $f(T) = a + \Phi \exp[ -dE/k T]$, immediately yields the gap energy $\dE$ and the prefactor $\Phi$ for $\krisc$ in equation \eqref{eq:krisc}.
In a similar manner, all of the primary constants ($k_1$, $k_3$, $\kisc$ and $\krisc$, or $\phi$ and $\dE$), can be obtained from simple curve fitting and combinations involving only the temperature dependent rate constants, as will be shown in the Section~\ref{sec:analysis}.

An alternative method for obtaining $\krisc$ and $\dE$ is based on delayed and prompt measurements of luminescent yields.\cite{Kirchhoff1983,Goushi2012}
Expanding the temperature dependence in equation \eqref{eq:delayedluminescence}, the general form for a delayed or prompt luminous yield is
\begin{equation}
  \label{eq:gatedyields}
  Y(t_{gate}) \sim \left[ a + b \e^{-(\dE/k_bT)} \right] \e^{-\left(\alpha + \beta \e^{-(\dE/k_bT)}\right) t_{gate}}
\end{equation}
where the measurement is made in a small window centered on a time $t_{gate}$,
and provided that $t_{gate}$ is well within either the prompt or delayed range as described above.
Note the second exponential in which appears both $t_{gate}$, and the exponential temperature dependence.
In other words, the expected relationship between yield and temperature seen in the first term is modified by the combined temperature and time dependence seen in the second exponential term. This presents additional complications for measurement and for data reduction and fitting, compared to the rate-constant-only method described above.

\begin{figure}[h!]
  \centering
  \includegraphics[scale=1.0]{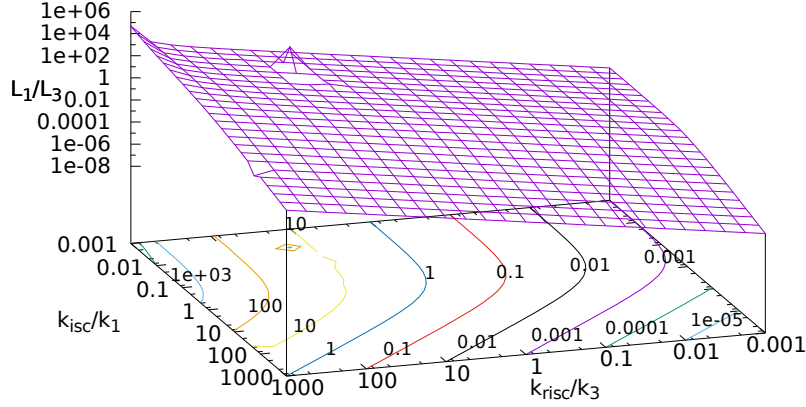}
  \caption{\label{fig:outputdelayedratio}
    Output ratio (single $L_1$ to triplet $L_3$) in delayed luminescence as a function of forward and reverse crossing rates in dimensionless units by scaling to spontaneous relaxation rates,
    $k_1 = 1, k_3 = 10^{-3}$.}
\end{figure}

In FIG.~\ref{fig:outputdelayedratio} the system of differential equations \eqref{sdecay},\eqref{tdecay} is integrated numerically to show the fluorescence to phosphorescence ratio $L_1/L_3$ evaluated at $t = 3/k_3$ as a function of $\kisc/k_1$ and $\krisc/k_3$ with $k_1 = 1, k_3 = 0.001$.
The result closely aligns with direct evaluation of equation \eqref{eq:tranratio}.
Comparing this result to FIG.~\ref{fig:outputratio}
we see that the delayed transient fluorescence/phosphorescence ratio is different in shape and has a larger range compared to the steady state ratio.

It is important to note that FIG.~\ref{fig:outputratio} and FIG.~\ref{fig:outputdelayedratio} display output ratios $L_1/L_3$.  The excited state population ratio $N_1/N_3$ is lower by a factor of $k_1/k_3$, which can be order $10^3$.  From FIG.~\ref{fig:outputratio}, the range for $N_1/N_3$ is from $1$ to $10^{-6}$.

This completes our model and closed form solutions for steady state and transient luminescence in a three level TADF system.

\section{\label{sec:analysis}Analysis of TADF Data}
Delayed fluorescence transients, as we have shown, are characterized by a fast (prompt) rate constant $r_1$ and a slow (delayed) rate constant $r_2$.
Here we describe a method for obtaining $k_3$, $k_1$, $\kisc$, $\krisc$ and $\dE$ from $r_1$ and $r_2$ measured at several temperatures,
and a second method by which $\dE$ and the prefactor $\Phi$ are obtained from $r_1 + r_2$.
We then apply the methods to data published for two TADF materials and calculate the fluorescence/phosphorescence and singlet/triplet ratios using equation \eqref{eq:steadyratio} and equation \eqref{eq:tranratio}.

\begin{figure}[h!]
  \subfigure[\ 4CzIPN, delayed]{
    \includegraphics[scale=0.38]{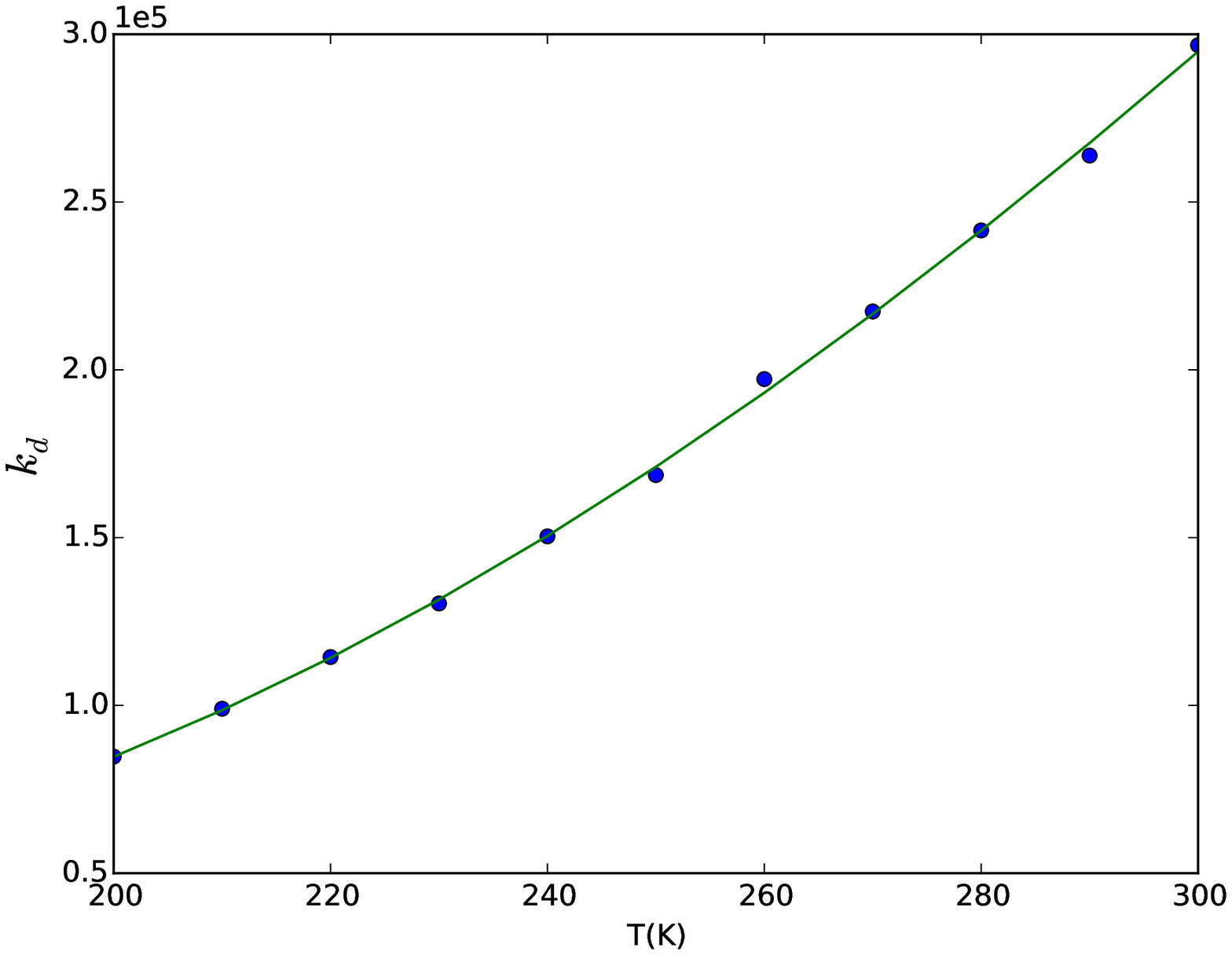}
    \label{fig:4Cr2fit}
  }
  \subfigure[\ m-MTDATA:t-Bu-PBD, delayed]{
    \includegraphics[scale=0.38]{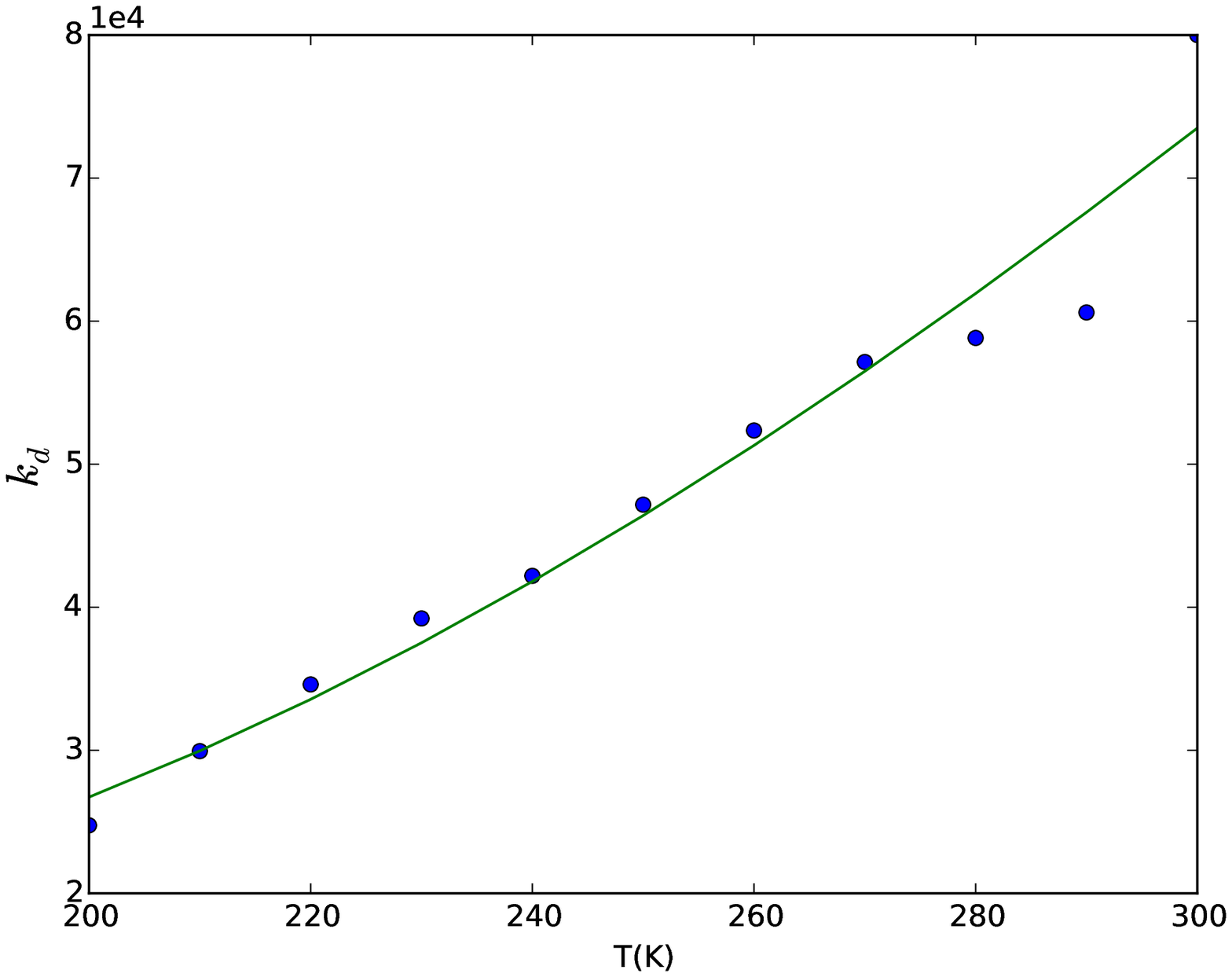}
    \label{fig:MTr2fit}
  }
  \subfigure[\ 4CzIPN, prompt]{
    \includegraphics[scale=0.38]{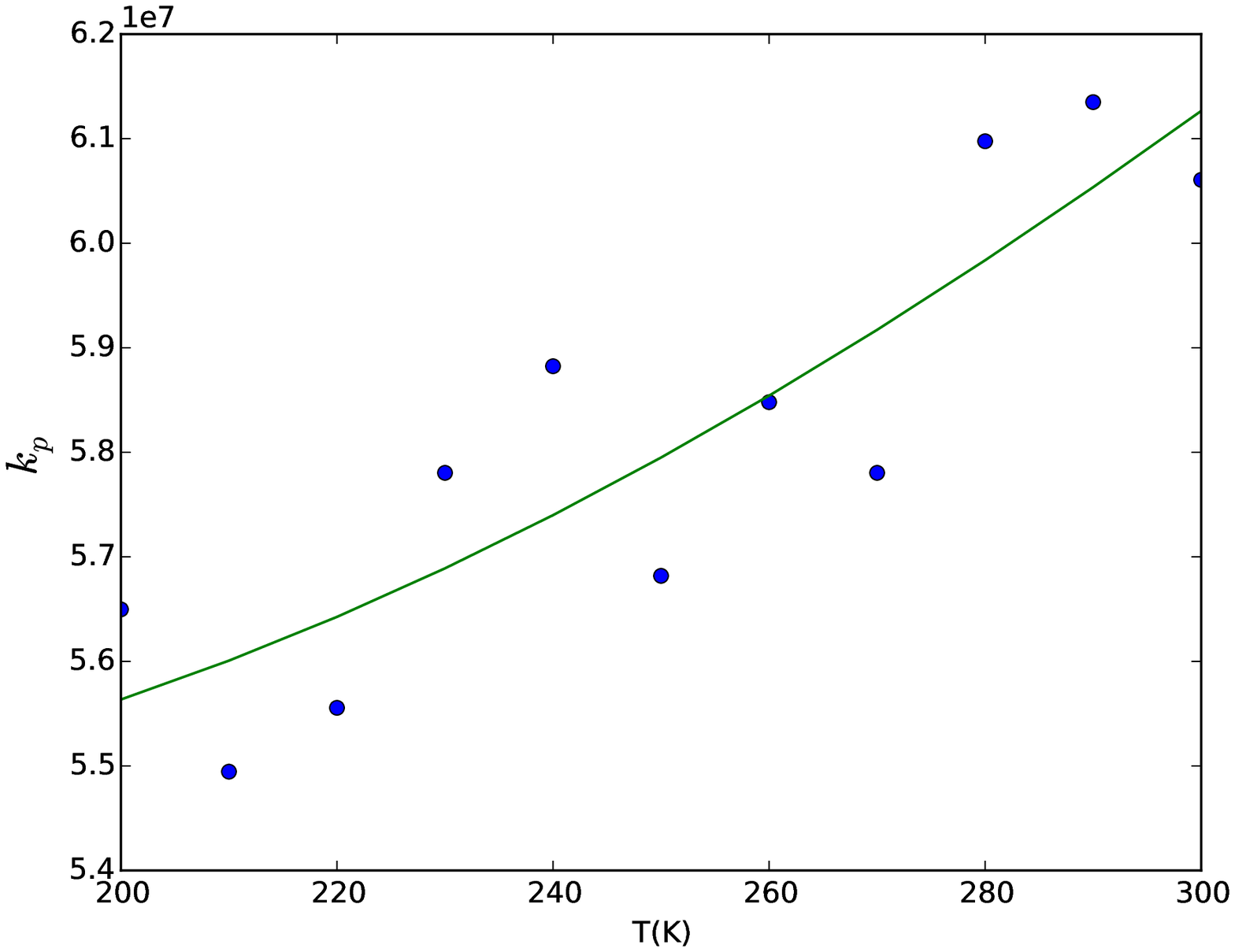}
    \label{fig:4Cr1fit}
  }
  \subfigure[\ m-MTDATA:t-Bu-PBD, prompt]{
    \includegraphics[scale=0.38]{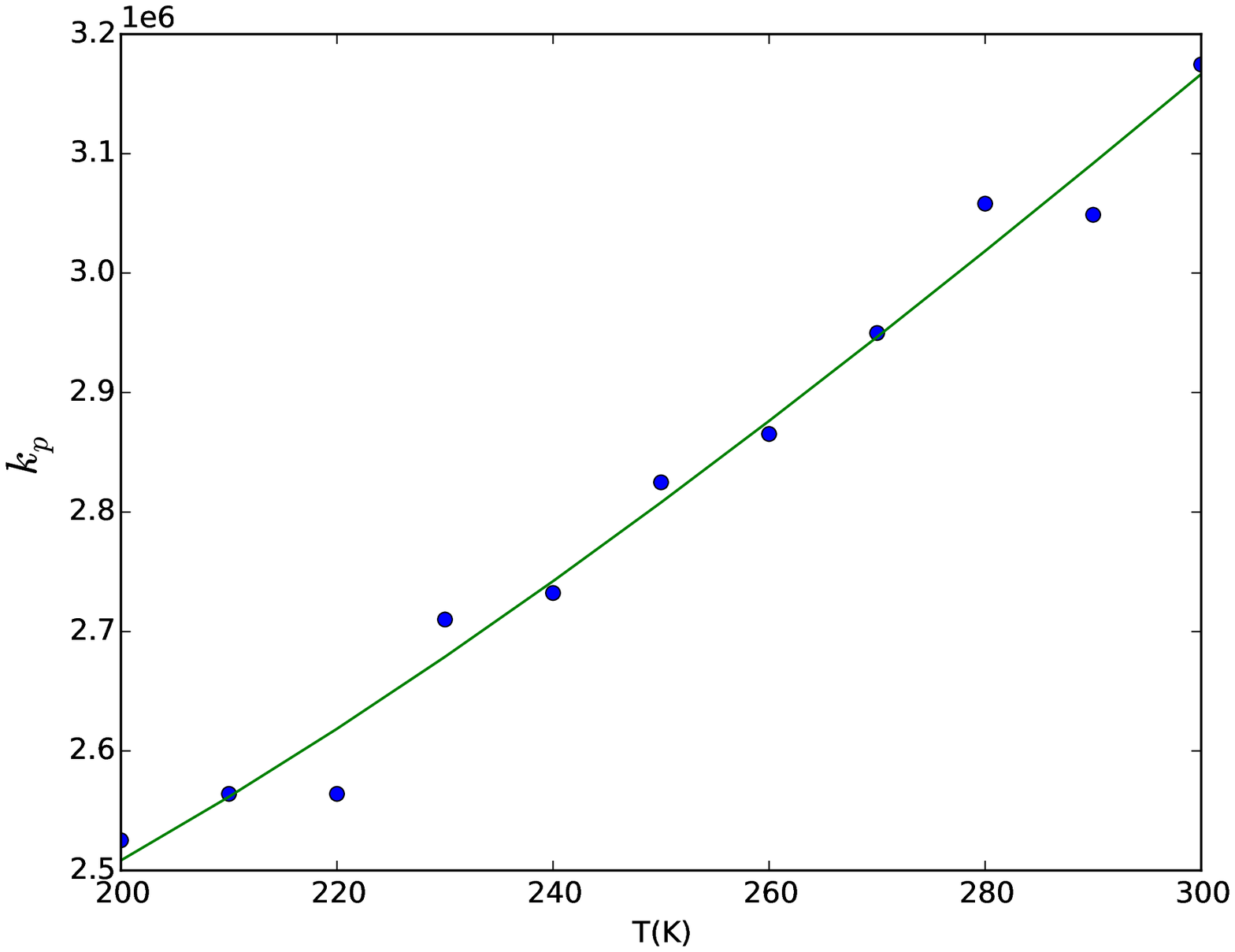}
    \label{fig:MTr1fit}
  }
  \caption{\label{fig:ratefits} Fit for delayed and prompt relaxation constants. Data are from supplemental materials provided by Goushi, et. al.\cite{Goushi2012} and Uoyama, et. al.\cite{Uoyama2012}.}
\end{figure}

We can write equation \eqref{eq:r2} for $r_2$, as
\begin{equation}
  -r_2 \approx k_3 + \left(1 - \frac{\kisc}{\kd}\right) \krisc
  \approx k_3 + \phi_3 \e^{-\dE/k_B T}
\end{equation}
provided that $\kd$ is nearly constant, or equivalently that $k_1 + \kisc - k_3 >> \krisc$ which will generally be true since $k_1 >> k_3$.
Fitting delayed decay rate data measured at several temperatures  (c.f. FIG.~\ref{fig:4Cr2fit} and \ref{fig:MTr2fit}), then gives us $k_3$ and $\dE$ along with the quantity $\phi_3$.

Similarly, we can write equation \eqref{eq:r1} for $r_1$, as
\begin{equation}
  -r_1 \approx k_1 + \kisc + \frac{\kisc}{\kd} \krisc = k_1 + \kisc + \phi_1 \e^{-\dE/k_B T}
\end{equation}
Fitting prompt decay rate data measured at several temperatures (c.f. FIG.~\ref{fig:4Cr1fit} and \ref{fig:MTr1fit}),
gives us the quantity $k_1 + \kisc$ and again $\dE$, along with the quantity $\phi_1$.

\begin{figure}[h!]
  \subfigure[\ 4CzIPN, $r_1 + r_2$]{
    \includegraphics[scale=0.38]{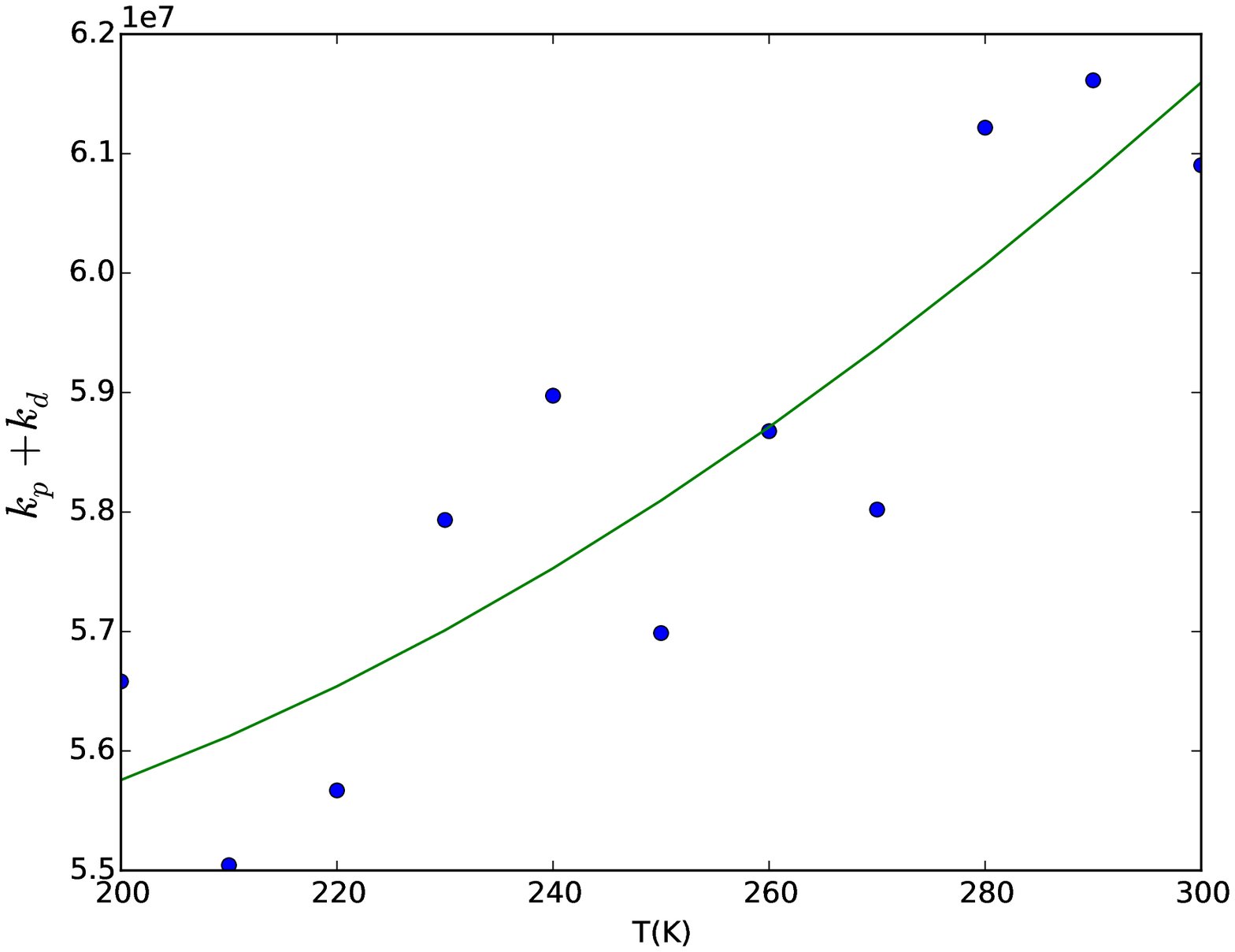}
    \label{fig:4Crsumfit}
  }
  \subfigure[\ m-MTDATA:t-Bu-PBD, $r_1 + r_2$]{
    \includegraphics[scale=0.38]{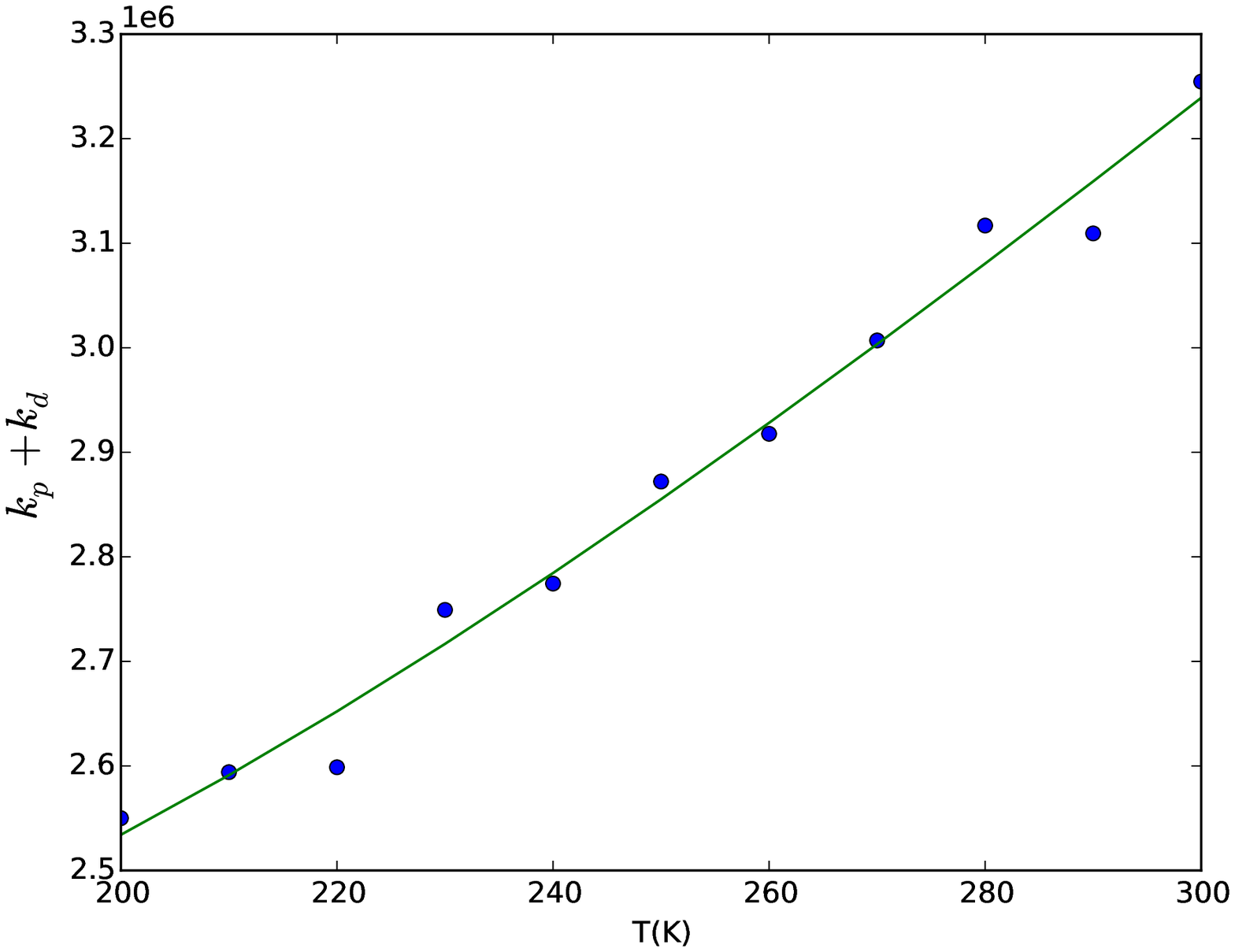}
    \label{fig:MTrsumfit}
  }
  \caption{\label{fig:rsumfits}
    Fit for combined delayed and prompt relaxation constants ($r_1 + r_2$).
    Data are from supplemental materials provided by Goushi, et. al.\cite{Goushi2012} and Uoyama, et. al.\cite{Uoyama2012}.
    The 4CzIPN fit exhibits scatter in the prompt relaxation time measurements compared to the reduced scatter in the m-MTDATA:t-Bu-PBD prompt relaxation time data.}
\end{figure}

Summing equations \eqref{eq:r1} and \eqref{eq:r2} we have
\begin{equation}
  \label{eq:rsum}
  r_1 + r_2 = -(k_1 + \kisc + k_3 + \krisc)
\end{equation}
and we can now obtain $\krisc$ by either of two methods. The first is that we simply add $r_1$ and $r_2$ measured at some temperature $T$, and subtract the already obtained quantities $k_1+\kisc$ and $k_3$. Alternatively, we can write equation \eqref{eq:rsum} as
\begin{equation}
  \label{eq:rsumfit}
  -(r_1 + r_2) = (k_1 + \kisc + k_3) + \Phi \e^{-\dE/k_b T}
\end{equation}
and then fit the temperature dependent quantity $r_1+r_2$ to obtain $\krisc$ explicitly in terms of its prefactor and gap energy (c.f. FIG.~\ref{fig:rsumfits}).

Then finally, we obtain $\kisc$, and $k_1$, from $\phi_1$ and $\phi_3$,
\begin{equation}
  \kisc = \kd \frac{\phi_1/\phi_3}{1 + \phi_1/\phi_3}
\end{equation}
Fluorescent and singlet ratios can then be calculated from equations \eqref{eq:steadyratio} and \eqref{eq:tranratio}.

Prompt and delayed rate constants as a function of temperature are provided in Uoyama,~et.~al.~(2012),\cite{Uoyama2012} and Goushi,~et.~al.~(2012),\cite{Goushi2012}, in supplemental materials.
Applying the above procedure to analyze the reported data, we obtain the values listed in Table~\ref{tbl:fits}.

\begin{table}[h!]
  \setlength{\tabcolsep}{0.25em} 
  \begin{tabular}{lcccccccc}
    Material & $k_1(10^6)$ & $\kisc(10^6)$ & $k_3 (10^3)$ & $\krisc^{(RT)}(10^6)$ & $\Phi(10^6)$ & $\Delta E$ & $L_1/L_3$ & $N_1/N_3$\\
    \hline
    4CzIPN            & 8.5 & 46  &29 & 6.7 & 200 &0.081 &\ 37 (36) &\ 0.13 (0.12)\\
    m-MTDATA:t-Bu-PBD & 1.2 & 1.1 &13 & 1.0 & 11 & 0.062 &\ 47 (43) &\ 0.53 (0.48)\\[0.2em]
  \end{tabular}
  \caption{\label{tbl:fits}
    Values in seconds and electron-volts obtained using data provided in Uoyama 2012 and Goushi 2012 (Adachi, et. al.).
    Fluorescence and singlet ratios in the delayed transient are shown in parenthesis.
  }
\end{table}

Examining the configuration of constants for the two materials, we see a large prefactor for the engineered material 4CzIPN along with a modestly larger gap energy compared to that in m-MTDATA:t-Bu-PBD, and find that the reverse crossing rates and population ratios are within an order of magnitude.
The modest singlet/triplet ratios produce large flourescence/phosphorescence ratios because of the large $k_1/k_3$ ratios.
High efficiency reported in OLEDS using these materials likely originates in competition between reverse crossing and losses in the triplet.
Thus, charge and self-quenching losses in a conventional OLED architecture, will eventually outpace reverse crossing and produce roll-off.

The $r_1$ and $r_2$ fits are shown in FIG.~\ref{fig:ratefits}. It might be noted that scatter is low in panels (a) and (b) where the rates are order $10^5/s$ and less, increases in panel (d) where the rates are order $10^6/s$ and increases again in panel (c) where the rates are $10^7/s$. Since the magnitude of the scatter (as well as on a percentage basis) is greater at the faster decay rate, the scatter might be related to gate timing as described above.
The effect of the data scatter on the calculated singlet/triplet population is about 10\%.

Uoyama et. al., for 4CzIPN report $\dE$ as 83 meV compared to our 81 meV,
$\krisc$ as 1.3\EE{6}/s at 300K, compared to our 6.7\EE{6}/s, and for $\kisc$ they use 4\EE{7} compared to our 4.6\EE{7}.
Goushi, et. al., for m-MTDATA:t-Bu-PBD, report $\dE$ as 50 meV compared to our 60 meV, $\krisc$ as 1.3\EE{5} compared to our 1.0\EE{6}, and for $\kisc$ they use 1.3\EE{5} compared to our 1.0\EE{6}.
Thus the values obtain in our analysis are similar to the previously reported values.

The earlier values were obtained by first calculating reverse crossing rates at each of a set of temperatures using the equation\cite{Goushi2012},
\begin{equation}
  \krisc = \frac{k_p k_d \phi_d}{\kisc \phi_p}
\end{equation}
where $k_d$ and $k_p$ are the delayed and prompt decay rates, $\phi_d$ and $\phi_p$ are the delayed and prompt yields, and $\kisc$ is the supplied forward crossing rate.
$\dE$ is then found as the slope in a log-linear fit to $1/k_B T$.
For 4CzIPN, $\krisc$ values are obtained with little scatter and $\dE$ is within a few percent of our value.
For m-MTDATA:t-Bu-PBD, there is considerable scatter in the $\krisc$ values and $\dE$ agrees less closely with our value.
As noted above in discussing FIG.~\ref{fig:ratefits}, there is more scatter in the prompt rate data for 4CzIPN compared to that of  m-MTDATA:t-Bu-PBD, and the $\dE$ value agrees more closely with ours.
Thus, in these two examples differences in $\dE$ seem to follow the $\krisc$ calculation in the earlier method, rather than the scatter in decay rates.


\section{Conclusions}

In this work we have obtained closed solutions for steady state and transient behaviors in TADF systems and described a procedure for extracting the primary relaxation and crossing rates, and the prefactor and gap energy from temperature dependent prompt and delayed decay rates.
We applied our method to data reported for two TADF materials and used the fitted values for the primary rate constants to calculate their fluorescence/phosphorescence and singlet/triplet population ratios.
The primary rate constants obtained in this manner are in reasonable agreement with reported values where available, and we are able to report the fluorescence/phosphorescence ratios 37:1 for 4CzIPN and 47:1 for m-MTDATA:t-Bu-PBD, along with their singlet/triplet population ratios 0.13 and 0.53.
It is hoped that the method will be useful in studies of TADF and OLEDS.

\end{document}